\documentclass[journal=apchd5,manuscript=article,layout=twocolumn]{achemso}

\usepackage[version=3]{mhchem} % Formula subscripts using \ce{}
\usepackage{color,soul}
\usepackage{setspace}

\author{Su-Peng Yu}
\affiliation{Time and Frequency Division, NIST, Boulder, Colorado, USA}
\alsoaffiliation{Department of Physics, University of Colorado, Boulder, Colorado, USA}
\email{supeng.yu@colorado.edu}

\author{Hojoong Jung}
\affiliation{Time and Frequency Division, NIST, Boulder, Colorado, USA}
\alsoaffiliation{Department of Physics, University of Colorado, Boulder, Colorado, USA}
\alsoaffiliation{Present address: Center for Quantum Information, Korea Institute of Science and Technology, Seoul, South Korea}

\author{Travis C. Briles}
\affiliation{Time and Frequency Division, NIST, Boulder, Colorado, USA}
\alsoaffiliation{Department of Physics, University of Colorado, Boulder, Colorado, USA}

\author{Kartik Srinivasan}
\affiliation{Microsystems and Nanotechnology Division, NIST, Gaithersburg, Maryland, USA}

\author{Scott B. Papp}
\affiliation{Time and Frequency Division, NIST, Boulder, Colorado, USA}
\alsoaffiliation{Department of Physics, University of Colorado, Boulder, Colorado, USA}

\title[Nanophotonic Kerr Comb]
  {Photonic-crystal-reflector nano-resonators for Kerr-frequency combs}

\begin{document}
\singlespacing

\begin{abstract}
\singlespacing We demonstrate Kerr-frequency-comb generation with nanofabricated Fabry-Perot resonators with photonic-crystal-reflector (PCR) end mirrors. The PCR group-velocity-dispersion (GVD) is engineered to counteract the strong normal GVD of a rectangular waveguide fabricated on a thin, 450 nm silicon nitride device layer. The reflectors provide the resonators with both the high optical quality factor and anomalous GVD required for Kerr-comb generation. We report design, fabrication, and characterization of devices in the 1550 nm wavelengths bands, including the GVD spectrum and quality factor. Kerr-comb generation is achieved by exciting the devices with a continuous-wave (CW) laser. The versatility of PCRs enables a general design principle and a material-independent device infrastructure for Kerr-nonlinear-resonator processes, opening new possibilities for manipulation of light. Visible and multi-spectral-band resonators appear to be natural extensions of the PCR approach. 
\end{abstract}

\section{Introduction}
\singlespacing
Optical-frequency combs provide revolutionary technologies for applications and research ranging from optical spectroscopy\cite{Coddington2016} and metrology\cite{Spencer2018} to optical communication\cite{Palomo2017}. In particular, nanofabricated frequency-comb systems would allow for wide-spread application of frequency-comb techniques, enabled by scalable fabrication at low cost. Kerr-microresonator frequency combs-- microcombs --realized with integrated photonics are also likely to be a potent propelling force for emerging technologies such as optical waveform generation\cite{Jiang2007,Cundiff2010}, ranging\cite{Trocha2018}, inertial navigation\cite{Li2017Brillouin}, and LIDAR applications \cite{Swann2006}. A variety of microcomb systems, with diverse materials and methods, are under active study.

Silicon nitride (Si$_3$N$_4$, hereafter SiN) is a thoroughly studied material for micro- and nano-photonic applications due to its low optical loss and compatibility with standard semiconductor manufacturing processes. SiN material is particularly advantageous for Kerr frequency comb generation in ring resonators due to high third-order nonlinearity\cite{Tan2010, Tien2010}, and has seen rapid development in recent years \cite{Pfeiffer2017}. The challenge for Kerr combs in the ring resonator geometry lies in the control of group-velocity dispersion (GVD, or simply dispersion). Typically, photonic waveguides with sub-wavelength cross-section demonstrate strong normal GVD associated with index guiding, and this inhibits phase-matching of four-wave mixing processes that are necessary for Kerr-comb generation. To maintain anomalous dispersion, a thick SiN device layer is often utilized, leading to challenges in fabrication due to etching thick layers and high materials stresses that lead to cracking and low device yield\cite{Pfeiffer2016}. Additionally, the bulk SiN material demonstrates strong normal dispersion at shorter wavelength ranges \cite{Luke2015}, resulting in challenges for Kerr-comb generation with visible light. Methods to alleviate such constraints are under active development, including exotic waveguide geometries\cite{Moille2018,Kim2017} and novel materials\cite{Rabiei2014}. %Do we have our TaO paper on Arxiv yet? Sadly, no.  But not sure tantala offers much direct benefit

In this article, we leverage sophisticated nanofabrication capabilities applicable to SiN photonic devices to create Fabry-Perot-type microresonators with photonic-crystal reflectors (PCR). Photonic crystals are periodic dielectric structures demonstrating unprecedented capabilities to engineer optical properties including group velocity\cite{Chung2018} and mode profile\cite{Yu2014,Feber2015}. The capability the PCR devices offer is that targeted optical characteristics are generated based on nanophotonic patterning independent of device layer thickness. A wider dynamic range of dispersion can be reached using this method to enable compensation of intrinsic dispersion contributions with a wider range of materials, hence enabling realization of Kerr comb generation in a broad range of integrated photonics platforms. For application in SiN devices proposed here, the requirement for a thick device layer can be alleviated, and the optical properties of such micro-cavities can be completely controlled by single-layer patterning using lithography. Here we demonstrate high optical quality factor resonators with engineered anomalous-dispersion capable of generating Kerr frequency combs, based on patterned SiN with a 450 nm thickness device layer.

The article will be presented as follows: first, we provide an intuitive concept for the PCR resonator Kerr combs, with numerical design methods and measurement plan; then, we report the fabrication and characterization, followed by comb generation experiment; finally, we propose possibilities made available by the PCR technology, including designs for a 532 nm pumped visible-band comb.
%By offloading the role of dispersion control to the PCRs,

\begin{figure}[h!]
\centering\includegraphics[width=\columnwidth]{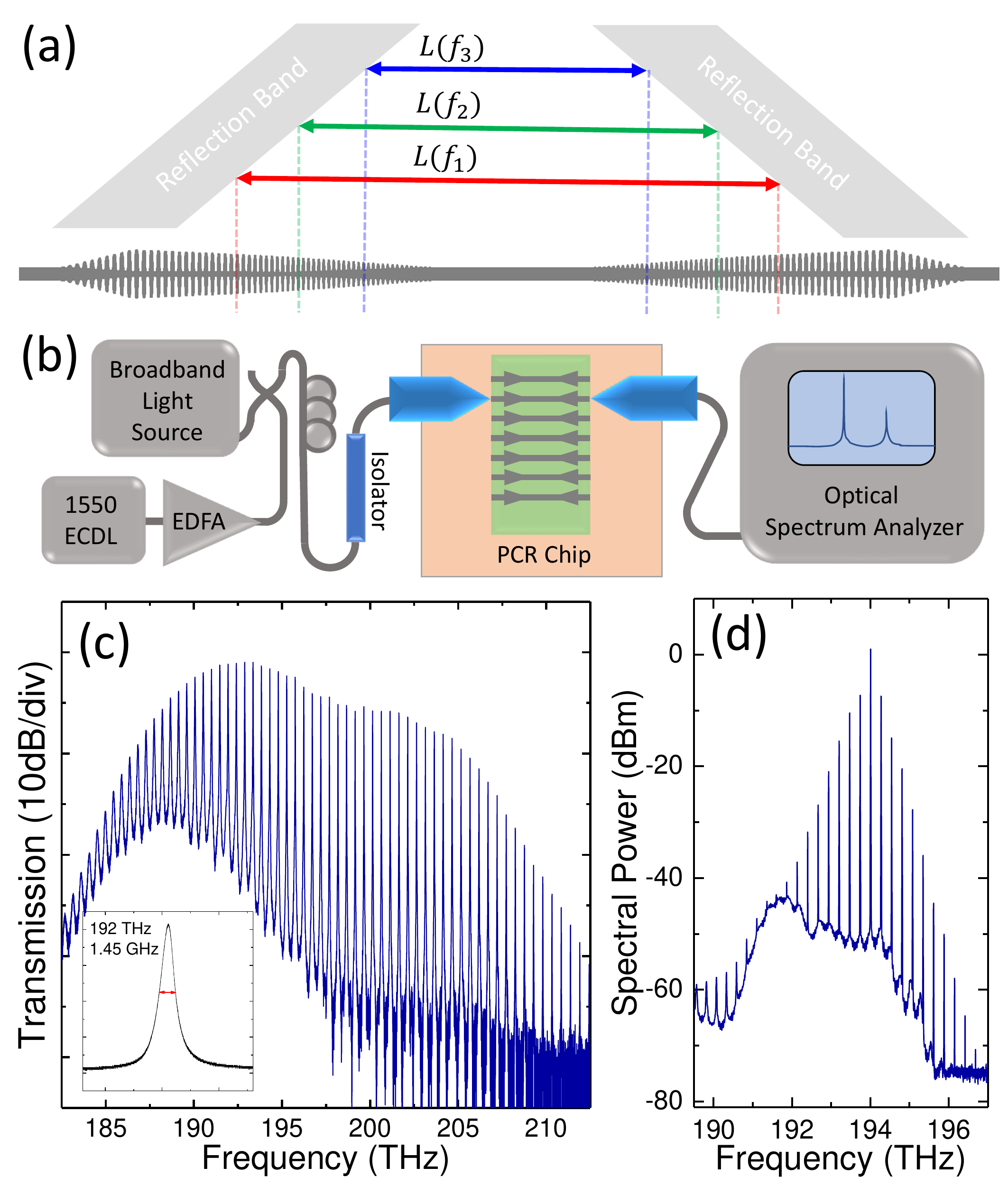}
\caption{Overview of PCR Kerr-frequency-comb generation. (a) Anomalous GVD engineering in PCRs, with the effective cavity length $L$ decreasing with increasing optical frequency $f_3 > f_2 > f_1$. (b) Optical testing system, shown with both broad-band and single-frequency light sources. (c) Broad-band transmission spectrum of a device, and (c inset) laser sweep across a resonance, with frequency calibrated against a 210 MHz Mach-Zehnder interferometer. (d) Frequency comb generation by pumping a device with amplified laser light.}
\label{Fig:Conceptual}
\end{figure}
%\hl{SP Figure comments: (a)  Can we make the resonator device picture bigger?  The $L_c$ is nice to understand the dispersion concept, however it is hard to see what the device is from this figure. Perhaps just expand the vertical (b)  Removing the PC and isolator would make the figure less busy.  Then the device would stand out better. (c) In the inset, remove the MZ fringe.  That's really hard to see.  Let's label the vertical axis as resonator transmission and label the units as dB.  (d)  Place the vertical axis tick markers and label on right side.}
%\textcolor{red}{(a) Moved the frequency domain plots not overlap with device to make it more visible. (b) Enlarged chip and drew cartoon of devices (c) removed fringes in inset. (d) Moved title to left. Still need to fix text size.}

\section{Design Concepts}

Figure \ref{Fig:Conceptual}(a) presents a conceptual picture of our PCR resonators and their use for Kerr-frequency-comb generation. The resonators are composed of two PCRs, and light is guided between them by a waveguide. The dispersion manifests in resonators in the form of frequency dependence of the free-spectral range (FSR) between adjacent resonances. The free-spectral range has the form $\textrm{FSR}=c/2\,n_gL$ in a Fabry-Perot resonator, where $n_g$ is the group index. In a Fabry-Perot-type resonator formed with sub-wavelength cross-section photonic waveguides, strong normal geometric dispersion arises from gradual concentration of optical field into the high-index waveguide region as the wavelength is reduced, therefore increasing $n_g$ of the guided optical mode. In such a normal-dispersion condition, FSR decreases with increasing optical frequency. To counteract decreasing FSR, the frequency-dependent effective cavity length $L$ can also be utilized as a engineerable parameter. We design PCRs to effectively reduce the cavity length $L$ with increasing optical frequency. Specifically, a PCR can be designed to reflect light at varying depth, with the highest optical frequencies reflecting at the shallowest depth. This is achieved by varying the local bandgap structures on the PCR (gray shaded areas in fig. \ref{Fig:Conceptual}(a)). The total GVD of a PCR resonator is a combination of the PCR contribution and the waveguide contribution, which scales directly with the waveguide length. Therefore, we may adjust the total dispersion simply by changing the length of the resonator. %, but this also changes the FSR.

%The changing in reflection position resulted in the effective cavity length to decrease with increasing optical frequency, hence increasing the FSR. The total group dispersion contribution of the waveguide section scales with its length, allowing desired total dispersion for the resonator to be tuned from normal to anomalous, simply by varying the length of the waveguide.

\subsection{Optical Characterization}

Figure \ref{Fig:Conceptual}(b) presents the experimental setup used to characterize both passive device properties, such as broadband on-resosance transmission (c) and observing Kerr comb formation (d). In particular, broadband transmission measurements offer a rapid testing procedure to rapidly explore the parameter space for PCR designs. We use an unseeded semiconductor optical amplifier as a broad-band light source, and a polarization controller and lensed fibers are used to launch light onto the silicon chip. The transmitted light is directly fed into an optical spectrum analyzer (OSA). The PCR resonators transmit light only on resonance, producing a low-background signal on the OSA even for high-Q devices with linewidth smaller than the resolution bandwidth of the OSA. Fitting the peak centers of the broad-band spectra allows for rapid extraction of device dispersion information. We are able to efficiently screen through devices to find ones with the desired reflection band placement and dispersion profile. Selected PCR resonators are tested in detail with tunable laser sweep for accurate determination of Q and coupling condition, shown in the inset of \ref{Fig:Conceptual}(c). We experiment with devices with proper dispersion and high Q factors to create an optical frequency comb from the PCR cavities. The tunable laser is amplified with an erbium-doped fiber amplifier (EDFA) to provide sufficient power to bring the PCR cavities above the comb generation threshold. An example PCR Kerr comb, analyzed by the OSA, is shown in Fig. \ref{Fig:Conceptual}(d).

\subsection{PCR design through finite-Element Simulations}

\begin{figure}[h!]
\centering\includegraphics[width=\columnwidth]{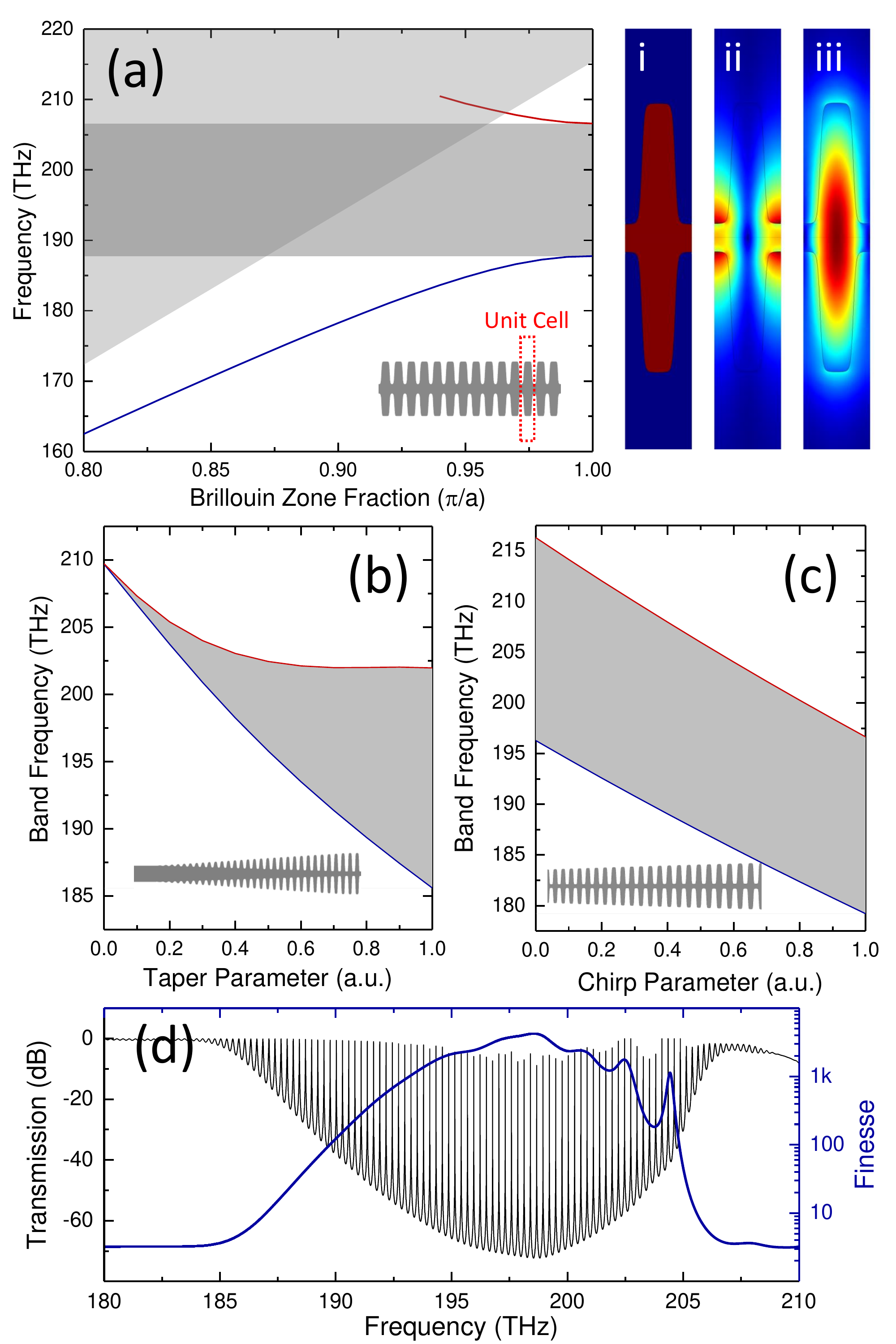}
\caption{(a) Band structure of the 1D photonic crystal, where $a$ is the lattice constant. At higher frequencies, the upper band (red) couples to free-propagating continuum and is no longer guided. The profiles of (i) nominal unit cell, (ii) upper and (iii) lower band electric field are also shown. (b) Local band structures of an adiabatic taper from ordinary waveguide to photonic crystal reflector. (c) Local band structure of a chirped photonic crystal reflector. (d) Calculated cavity finesse and simulated transmission using an 1D model with reflection coefficient from FEM results and a designated waveguide length.}
\label{Fig:BandStructure}
\end{figure}

Here we present the design procedure for PCR resonators. There are two primary requirements to be satisfied by the PCRs. First, the PCR must show sufficiently high reflectivity in the design wavelength ranges, characterized by the finesse $F = \frac{FSR}{FWHM} \simeq \frac{\pi}{1-R}$, where $R$ is the reflectivity of the PhC reflector. Second, the PCR must generate anomalous group-delay dispersion (GDD) in the target wavelength ranges. For ease of comparison, the GDD will be converted to an effective GVD over the in-resonator waveguide by $\text{GVD = GDD} / L$. The PCRs are composed of sub-wavelength unit cells, whose geometry determines the local reflection band of the crystal, shown in \ref{Fig:BandStructure}(a). A PCR is then built from a stack of continuously varying cells. Finite-element method (FEM) tools are employed to explore the parameter space of unit-cell geometry, and the reflection and transmission characteristics of a stack of multiple cells.

Fig. \ref{Fig:BandStructure} presents our design process for PCRs to satisfy the requirements for Kerr-comb generation. A typical unit cell dispersion curve is plotted for the 1D Brillouin zone in Fig. \ref{Fig:BandStructure}(a), with its geometry and electric field patterns shown in panel (i-iii). Strong reflection is achieved by creating a photonic bandgap, a frequency range where no guided propagation mode exists. The bandwidth and center of the bandgap can be rapidly simulated with a FEM eigenfrequency solver in full 3D, since the simulation volume contains only one unit-cell. We rapidly map the bandgap placement as function of unit cell geometries using this method. The unit cells parameters are then used as components to design reflectors made from a stack of varying cells. A typical reflector begins as an unpatterned waveguide, the photonic crystal is introduced by adiabatic onset of modulation to the local width of the waveguide. For simplicity, we use a simple linear tapering scheme for the adiabatic onset, where the parameters such as the waveguide width, modulation amplitude, and lattice constant are varied linearly with the tapering parameter $\eta$:
$$\vec{v}(\eta) = \vec{v}_0 + \eta (\vec{v}_1-\vec{v}_0)$$
where $\vec{v}$ represents the list of unit cell geometry parameters. The local bandgap of such a taper is shown in Fig. \ref{Fig:BandStructure}(b), where the unit cell continuously transforms from a waveguide with width $=750$ nm and lattice constant $=430$ nm ($\eta=0$) to the nominal cell of minimum width $=200$ nm, lattice constant $=480$ nm, and modulation peak amplitude of 875 nm ($\eta=1$). The geometry is designed so that high optical frequencies reflect at shallower depth in the taper, and also to ensure the absence of localized resonance modes inside the reflector itself. The latter meaning no frequencies pass through two disjoint sections of bandgap on the tapering profile. We also design PCRs with chirped nominal parameters. The scale invariance of Maxwell's equations indicate that a change in reflection wavelength range $\lambda \rightarrow \xi \lambda$ can be achieved by scaling all geometry parameters by $\vec{v} \rightarrow \xi \vec{v}$. In practice, this is carried out for all parameters except the device layer thickness. This chirping scheme enables sweeping of bandgaps over a larger total bandwidth. The local bandgap of this chirped mirror is plotted in Fig. \ref{Fig:BandStructure}(c). The desired anomalous dispersion engineering can be achieved by either using the adiabatic tapering or chirping. We were able to achieve higher finesse with the former, while broader bandwidth with the later.

With a design in hand, the sweeping trajectory in the parameter space is realized with a finite number of sites, and the resulting stack geometry is simulated in 3D to extract the complex transmission and reflection coefficients of the PCR. With sufficiently slow sweeping of geometry, the scattering loss due to group index and mode profile mismatch can be adiabatically suppressed\cite{Oskooi2012}. A typical simulated finesse curve and an expected transmission spectrum assuming a Fabry-Perot-like 1D resonator formed from such PCRs are plotted in Fig. \ref{Fig:BandStructure}(d), showing a series of resonances with strong suppression of off-resonance transmission in the target bandwidth from 190 to 200 THz. %, while transmitting efficiently outside this range.

\section{PCR-resonator nanofabrication}

\begin{figure}[h!]
\centering\includegraphics[width=\columnwidth]{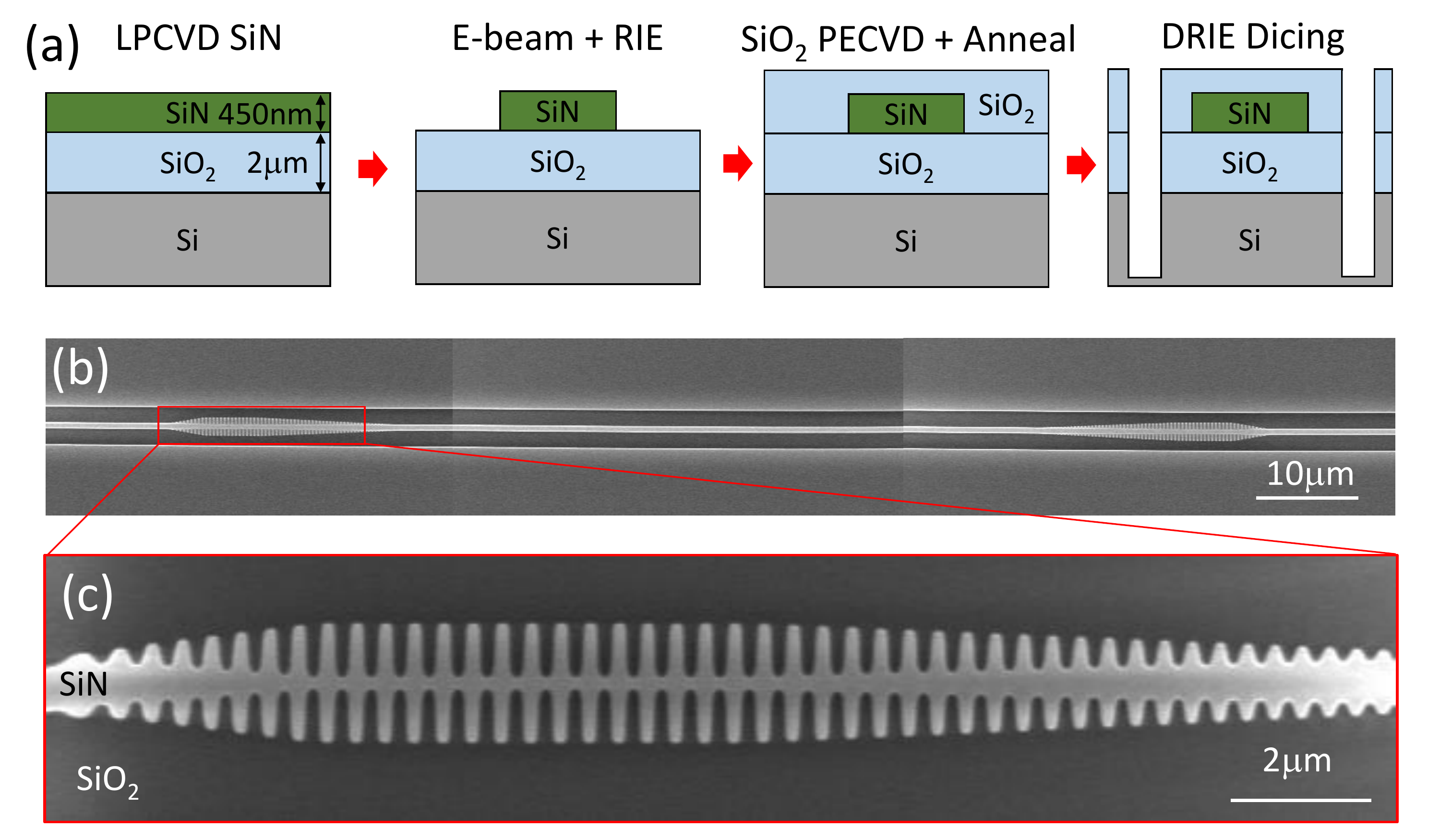}
\caption{(a) Fabrication process steps include e-beam lithography, pattern transfer by RIE, cladding using PECVD oxide, and chip separation with deep RIE. (b) A stitched-together full view of a resonator, and (c) zoom-in image of a photonic crystal reflector using SEM.}
\label{Fig:Fabrication}
\end{figure}

We fabricate PCR resonators by the process flow in \ref{Fig:Fabrication}(a). We acquire silicon wafers with thermally grown silicon dioxide for a cladding layer, and the silicon nitride device layer is grown with low-pressure chemical vapor deposition (LPCVD). We choose a device layer thickness of 450 nm to specifically allow for conventional, single-step LPCVD processing and subsequent nanofabrication without the need for stress-relieving patterns\cite{Luke2013, Pfeiffer2016}, annealing steps between multiple LPCVD runs\cite{Luke2015}, or other steps necessary to process thick SiN films. The devices are patterned using electron-beam lithography (EBL), and transferred into the device layer using a fluorine-based reactive ion etching (RIE) process. The patterns for the EBL step are adjusted empirically to compensate for dimension changes during pattern transfer, enabling us to fabricate devices in good agreement with the design geometries. We apply a top cladding layer of silicon dioxide, using a plasma-enhanced chemical vapor deposition (PECVD) process, as fully oxide-clad devices enable more efficient input power coupling using lens fibers. Top cladding also makes the devices more robust against contamination. We perform a high-temperature annealing step of 900 $ ^\circ$C for three hours to improve the optical quality of the PECVD oxide after deposition, and we observed up to a factor of three reduction of the optical absorption. % This number is from the JILA 'mysterious' Run0 annealing. Unfortunately I don't have before-after measurement of this batch. Unfortunately the annealing test results for PECVD coated chips were very limited.
To separate chips from the fabrication wafer, we use deep reactive ion etching, which also enable us to fabricate tapering bus waveguides terminating at the chip facets for high fiber-to-chip coupling efficiency. Overall, we fabricate chips with hundreds of PCR resonator devices, and lensed fiber coupling to the chip devices offers 3.5 dB insertion loss per facet.

\section{PCR-resonator GVD design and characterization}

\begin{figure}[h!]
\centering\includegraphics[width=\columnwidth]{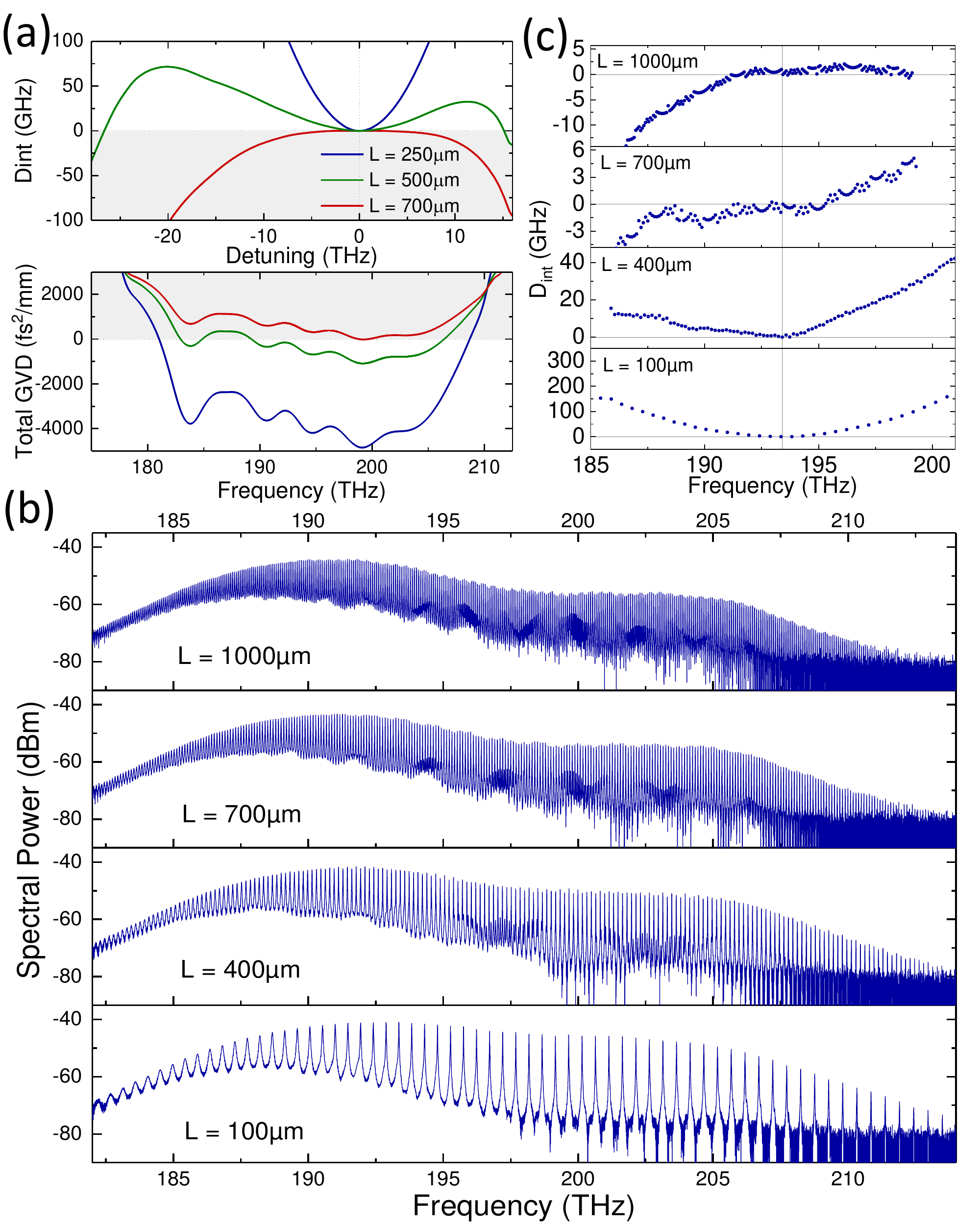}
\caption{(a) Calculated $D_{int}$ and total dispersion for cavities with chirped photonic crystal reflector and 500 nm width waveguide of varying length. (b) Measured broad-band transmission spectra of chirped reflector and 750 nm width waveguides of varying length, where the FSR are 70, 98, 184, and 484 GHz from panel one (top) through four (bottom), and (c) calculated $D_{int}$ from resonance frequencies demonstrating transition from anomalous to normal dispersion.}
\label{Fig:DispersionEngineering}
\end{figure}

We have carried out a detailed set of simulations to demonstrate GVD engineering in our PCR resonators. With the thin SiN device layer in this study, the total effective GVD is a competition between the normal dispersion of the waveguide and the anomalous effective dispersion of the reflectors. As the length of the resonator is increased, the total effective dispersion continuously varies from strongly anomalous from the reflectors to normal dispersion of the waveguides.  Here we engineer the anomalous PCR dispersion using the aforementioned tapering and chirping methods.  Following the convention mentioned previously, the GDD of the PCR will be averaged over the in-resonator waveguide length for the following discussion. A set of simulated traces demonstrating such dispersion change is plotted in Fig. \ref{Fig:DispersionEngineering}(a). Here, the design goal is to form a pocket of controlled anomalous dispersion around the Telecom C band. The resonator length is swept to demonstrate the dispersion balancing, where a near-zero dispersion is predicted at a length of 700 $\mu$m.

We can systematically vary the dispersion setting by the tapering condition-- with a longer taper resulting in larger anomalous dispersion. Our simulations indicate that a range of dispersion is achievable by this method, although constraints exist from the desire to also achieve high resonator finesse and efficient resonator coupling. Specifically, a taper that is too short in length is no longer adiabatic and causes scattering loss and reduces finesse, while one that is too long results in a reflector with insufficient transmission, hence under-coupling the resonator. For future designs, the restriction associated with insufficient PCR transmission can potentially be alleviated by evanescent coupling the bus waveguide directly to the in-resonator waveguide in a manner similar to the ring resonator case. The strength of normal dispersion of the waveguide section depends on the waveguide width, which can also be varied as a design parameter. %Therefore we can use this as a design parameter subject to considerations for small effective mode area to facilitate Kerr-nonlinear processes and low propagation loss.

%The total GVD of a PCR resonator results from the sum of the GDD of the PCRs and the GVD of the waveguide that forms the resonator.
%In Fig. 4, we explore the considerations related to PCR-resonator GVD design, and we characterize GVD in fabricated devices.
%It may be helpful to compare PCR resonator GVD with more traditional ring-resonator devices. In this case, we can convert PCR GDD
%The GDD from the reflectors are impinged onto the cavity twice per round-trip. For ease of comparison to waveguides and ring resonators, we will convert such GDD to an effective GVD simply by dividing the length of the resonator.

To put PCR-resonator dispersion engineering into practice, we fabricated a series of devices that feature a systematic GVD variation; see Fig. \ref{Fig:DispersionEngineering}(b,c). In particular, we design sweeps in device length to demonstrate the counter-balancing of waveguide and reflector dispersion. Transmission measurements using a broadband source, here a supercontinuum source \cite{lamb2018optical} for better measurement bandwidth, allow us to characterize the GVD in detail. The reflecting PCR resonators only transmit light on resonance, therefore the transmission as measured by the OSA is comprised of a set of narrow peaks; the resulting spectra are shown in Fig. \ref{Fig:DispersionEngineering}(b). We identify the frequency of the resonances by finding the peaks of the transmitted power, then calculate the resonator dispersion profiles $D_{int}$, shown in \ref{Fig:DispersionEngineering}(c). The dispersion profiles demonstrate the predicted gradual transition from strong anomalous dispersion for the \textit{L} = 100 $\mu$m to normal dispersion for the longest \textit{L} = 1000 $\mu$m case, with the zero-dispersion occurring near \textit{L} = 700 $\mu$m.

%\section{%Measurement Results}

\subsection{PCR-resonator Kerr-comb generation}

\begin{figure}[h!]
\centering\includegraphics[width=\columnwidth]{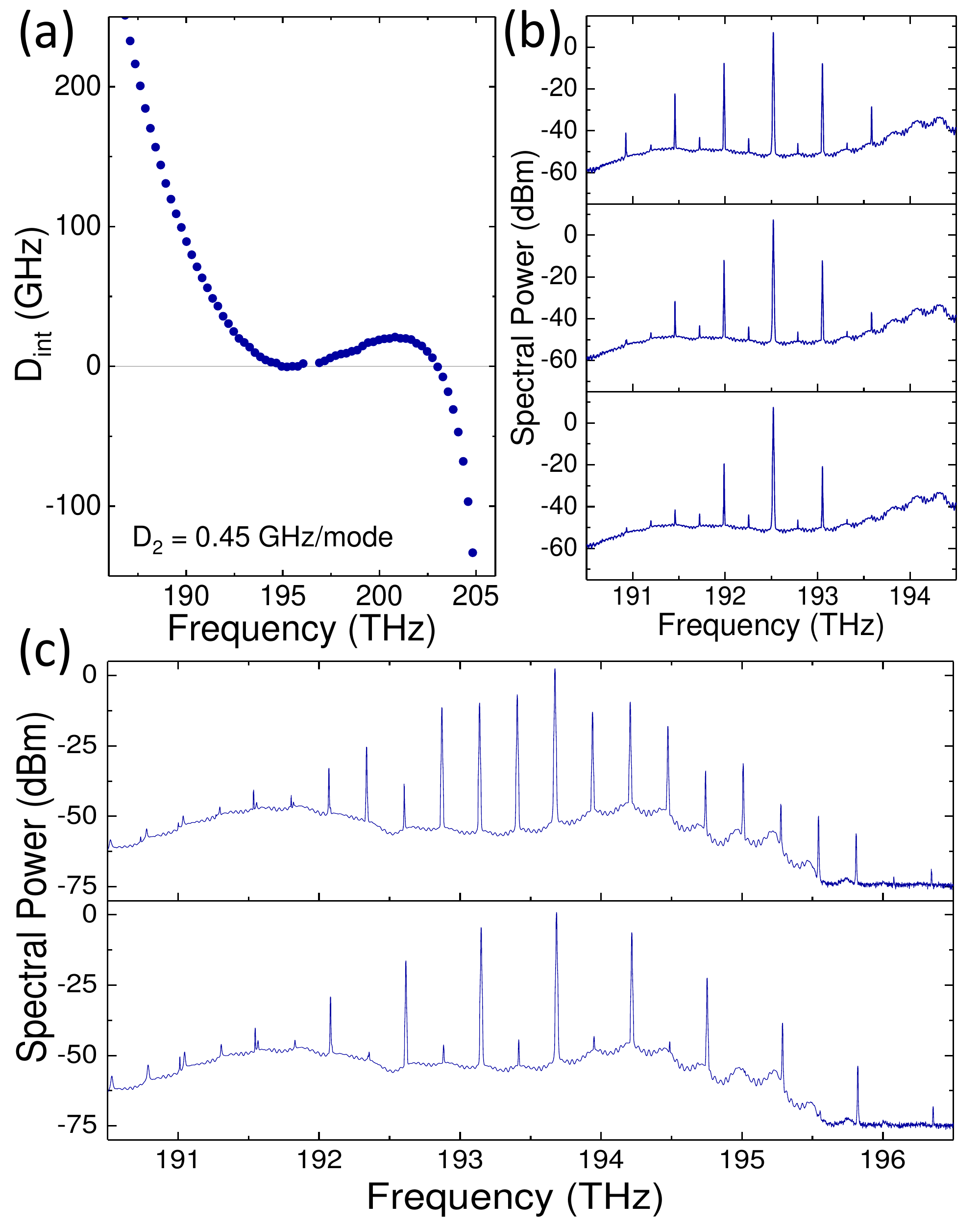}
\caption{(a) The calculated integrated dispersion $D_\text{int}$ relative to the pumped mode at 1550 nm. Here the device reflection band is chosen to be off-center from the pump to achieve desired coupling condition, leading to the asymmetric shape in $D_\text{int}$. (b) The onset of primary comb, and (c) the filling to a full comb with increasing pump detuning. We note that traces in (b) and (c) were measured from two different devices.}
\label{Fig:CombGeneration}
\end{figure}

A demonstration of Kerr-frequency combs generation with PCR resonators is shown in Fig. \ref{Fig:CombGeneration}. We selected a device length of $L_c=250 \, \mu$m, which is within the range described in the previous section. The loaded quality factor of the device is $\geq 10^5$. The specific dispersion parameters of the device are characterized by $D_2=-c/n \cdot \beta_2 \cdot {FSR}^2 \approx 0.45$ GHz/mode, where $\beta_2$ is the second-order GVD, and the integrated dispersion $D_\text{int}= \frac{D_2}{2}\cdot m^2+O(m^3)$, where $m$ is the mode index; see Fig. \ref{Fig:CombGeneration}(a). We amplify the CW pump laser in an EDFA and couple up to 550 mW onto the chip incident on the first PCR. As the pump laser is tuned into the cavity resonance starting on the blue detuned side, we observe characteristic thermo-optic bistability and self-locking of the resonator onto the pump laser allowing for a large buildup of intracavity power. Given sufficient power, the parameteric oscillation threshold is reached, and we observe the formation of a Kerr comb in the modulation instability regime excited from vacuum.

%The expected comb line generation, broadening, and filling-in are demonstrated in the transmitted light spectra, shown in Fig. \ref{Fig:CombGeneration} (c) and (d).%at a power of \hl{XX} mW, signal and idler sidebands of the pump laser appear. --> Need to re-measure to get

%\hl{Su-Peng:  to decide how much light is coupled to the resonator, I would suggest taking into account all losses, eg chip coupling and reduced coupling due to mirror finesse.} \textcolor{blue}{I think the actual in-waveguide power for these measurements are around 1W = 5W EDFA x 50pct delivery efficiency x 3.5dB loss at input facet. The issue with the past data was that I kept the EDFA at high power and was seaching across many devices to look for any device that makes something, hence I don't really have a precise number for threshold at this moment. Now that we know which devices work, it might be the easiest if we can spend an afternoon re-taking the comb generation data with the 5W fiber-coupled amplifier? This way we can include ASE filter too to make Fig5 look better.} 

%(a) Measured dispersion $D_2$ of a comb-generation testing device. The apparent discrete steps in (a) is likely an artifact associated with digitization of our data. 

Fig. \ref{Fig:CombGeneration}(b,c) show the evolution of comb generation, operationally as we increase in-resonator power by tuning the CW laser onto resonance. We observe Turing-pattern generation-- so-called primary comb --in modes $\pm$2 FSRs away from the pump mode. Harmonics of the primary comb lines become populated with increasing in-resonator power, shown in Fig. \ref{Fig:CombGeneration}(b). We also observed filling-in of the comb lines to form a full 1-FSR (268 GHz) comb, given sufficient power, shown in Fig. \ref{Fig:CombGeneration}(c). The elevated background noise level in this figure is the amplified spontaneous emission noise from the amplifier used with the CW laser. The comb generation unambiguously verifies the creation of anomalous dispersion in the PCR resonators. The kinds of Turing patterns that we observe are the fundamental process of any Kerr-microresonator. Our experiments show the opportunity to systematically vary the optical-mode spacing of Turing patterns, which offers functionality to generate millimeter-wave electronic signals through photodetection. Moreover, access to Turing patterns offer the possibility for optical-parametric amplification processes of external fields injected to the PCR resonator and related optical-wavelength translation processes. In future device fabrication iterations, we plan to target higher $Q$ factors that can both lead to lower threshold and potentially access to Kerr-soliton generation.

%As the pump laser frequency is swept closer into resonance with the PCR cavities, the primary comb populates multiples of twice the cavity FSR, as shown in Fig. \ref{Fig:CombGeneration}(b).

%optical frequency comb in the PCR cavities, we selected a device length of 250 $\mu$m, with a loaded quality factor of $\geq 10^5$, while providing a total dispersion in a reasonable range. The D2 values and a measured second-order dispersion $D_{int}$ for one such device is plotted in Fig. \ref{Fig:CombGeneration}(a) and (b).  We pump the cavity with an ECDL laser amplified with an EDFA. An input power of up to 3.2W can be delivered to the lens fiber, which, with an insertion loss of 3.5 dB per facet for these devices, corresponds to 1.4W power in the waveguides, sufficient to reach comb generation threshold for the Q factors available to these devices. 
 %\subsection{Frequency Comb Generation}

\section{Future Prospects}
\begin{figure}[h!]
\centering\includegraphics[width=\columnwidth]{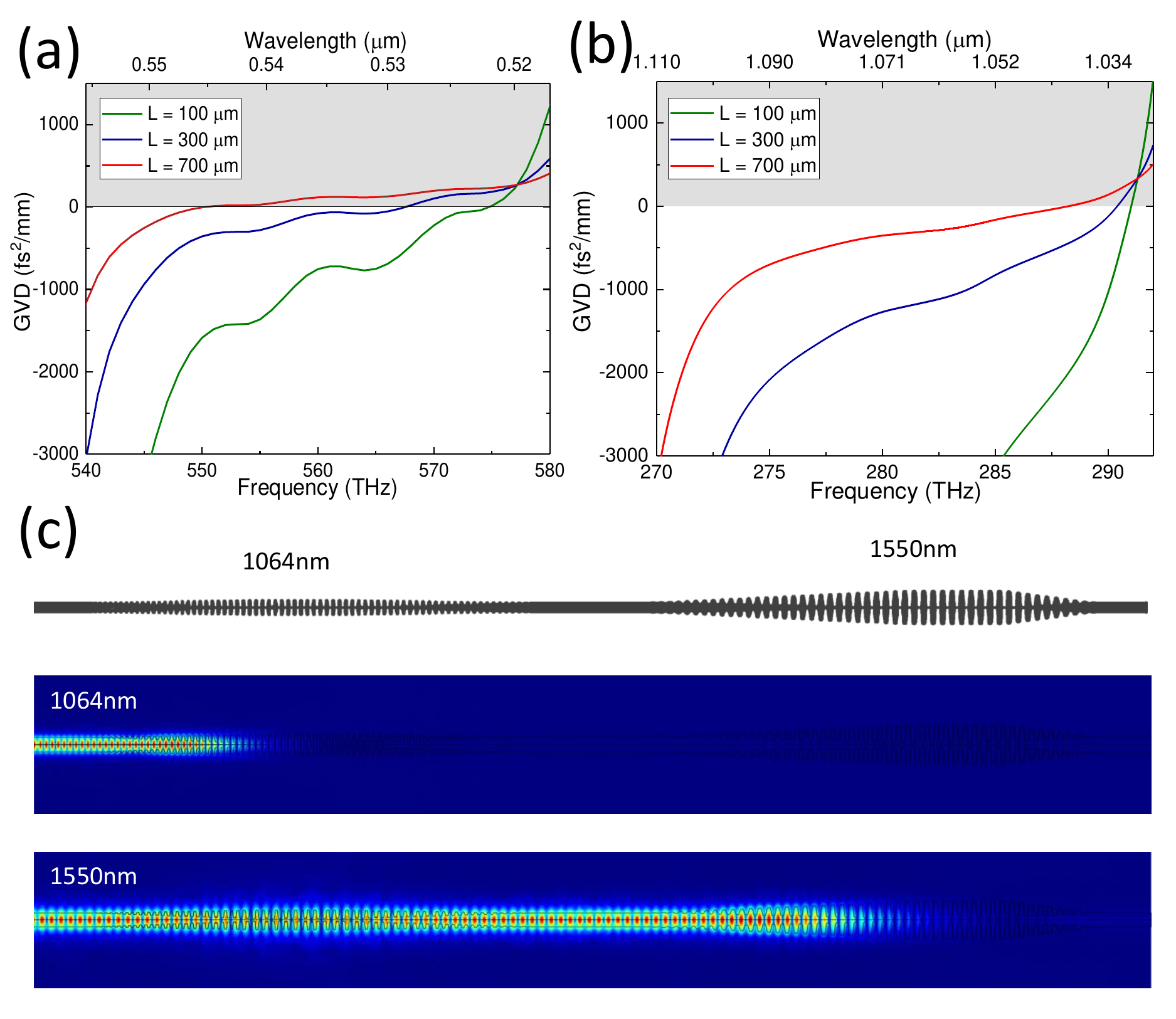}
\caption{(a) Total dispersion for photonic resonators optimized for anomalous dispersion at 532 nm wavelength, and (b) 1064 nm wavelength. (c) Illustration and simulated electric field profiles for 1064 nm to 1550 nm cascaded photonic crystal reflectors.}
\label{Fig:FutureProspects}
\end{figure}

We would like to emphasize that a significant advantage of PCR-based resonators is the decoupling of device layer configuration from dispersion engineering. To demonstrate this capability, we utilize the same design principle as the 1550 nm band devices described above to create anomalous dispersion for multiple wavelength bands. An interesting case is shown in Fig. \ref{Fig:FutureProspects}(a,b), where anomalous dispersion is created around 532 nm and 1064 nm wavelength ranges, using a shared SiN device layer thickness of 450 nm. The strong anomalous dispersion of the PCR is sufficient to balance the typically strong normal dispersion of materials in the visible ranges. The capability to engineer dispersion profiles independent of device layer enables integration of multiple wavelength band devices onto a same chip. Taking this concept further, we point out that the PCR is transparent to frequencies below its bandgap, making it possible to build cascaded reflectors like the case shown in \ref{Fig:FutureProspects}(c). Multi-frequency cascaded resonators as such can potentially provide enhancement for second- or third-harmonics generation, or to facilitate interlocked combs between different wavelength ranges \cite{Moille2018}. The PCR resonators provide flexibility in dispersion control beyond conventional ring- and disc-resonators, and makes available dispersion parameter ranges previously difficult to reach.

\section{Conclusion}

We have presented on-chip Fabry-Perot-type resonators with tailored dispersion based on PCRs, providing a pathway toward integrated photonics frequency comb generation. Conceptually, the PCRs are the microscopic analogy of chirped dielectric stack mirrors, capable of producing anomalous dispersion based on their geometry, independent of the device layer configuration. We designed and fabricated PCR resonators based on a SiN device layer and standard fabrication techniques. Using a broad-band light source and an OSA, we were able to directly verify the dispersion profile of such resonators. Anomalous dispersion of varying strength is created in a 450 nm thickness device layer, which is challenging to achieve with conventional ring resonators. We also demonstrated frequency comb generation in these devices. The PCR cavities provide versatile capabilities in dispersion engineering, and should enable construction of on-chip frequency comb sources for a wide range of design wavelengths.

%Need to cite something so achemso doesn't crash; See \cite{Spencer2018}.


%%%%%%%%%%%%%%%%%%%%%%%%%%%%%%%%%%%%%%%%%%%%%%%%%%%%%%%%%%%%%%%%%%%%%
\begin{acknowledgement}

Funding provided by DARPA DODOS, Air Force Office of Scientific Research (FA9550-16-1-0016), NIST, and UMD/NIST-CNST Cooperative Agreement (70NANB10H193). We acknowledge the Boulder Microfabrication Facility, where the devices were fabricated. We thank Nima Nader and Jeff Chiles for helpful suggestions, and Tara E. Drake and Jizhao Zang for a careful reading of the manuscript. This work is a contribution of the U.S. Government and is not subject to copyright. Mention of specific companies or trade names is for scientific communication only, and does not constitute an endorsement by NIST.

\end{acknowledgement}

% \begin{suppinfo}

% This will usually read something like: ``Experimental procedures and characterization data for all new compounds. The class will automatically add a sentence pointing to the information on-line:

% \end{suppinfo}

\bibliography{PhCCombBib}

% Some journals require a graphical entry for the Table of Contents.
% This should be laid out ``print ready'' so that the sizing of the
% The surrounding frame is 9\,cm by 3.5\,cm, which is the maximum
% permitted for  \emph{Journal of the American Chemical Society}

\makeatletter
\setlength\acs@tocentry@height{1.85in}
\setlength\acs@tocentry@width{1.7in}
\makeatother

\begin{tocentry}
\centering\includegraphics[width=1.85in,height=1.7in]{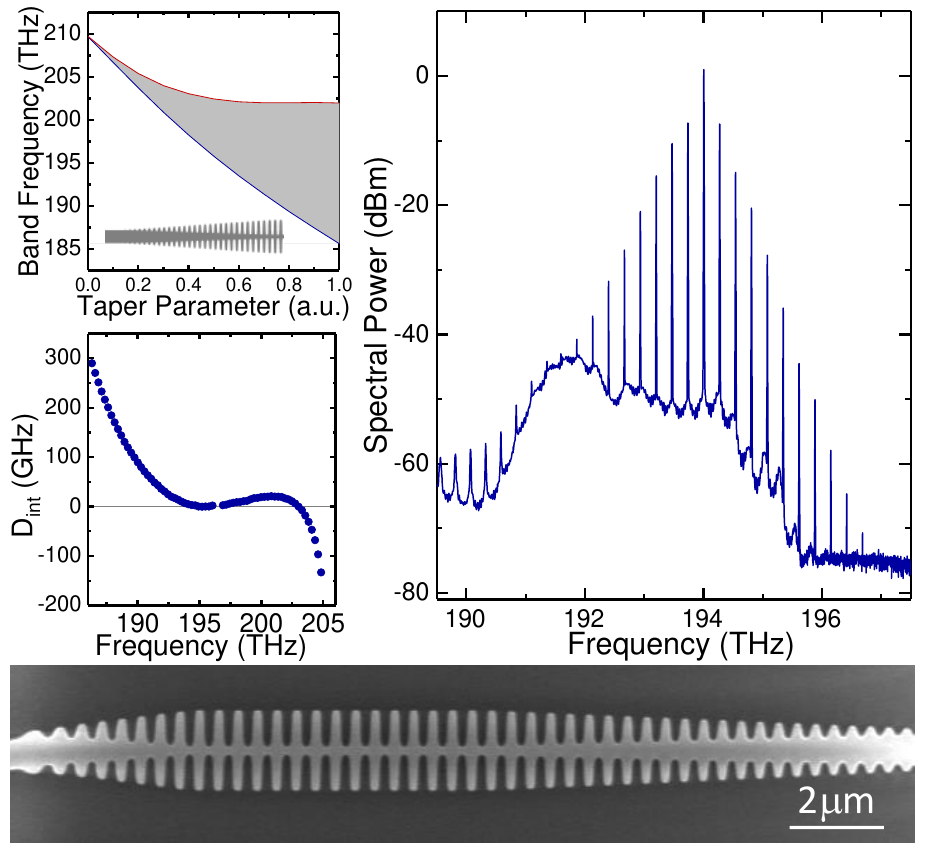}
\label{Fig:ToC}
\end{tocentry}

\end{document}